\documentclass[11pt]{article}
\usepackage{moriond,epsfig}

\def\MyPhoto{\psfig{figure=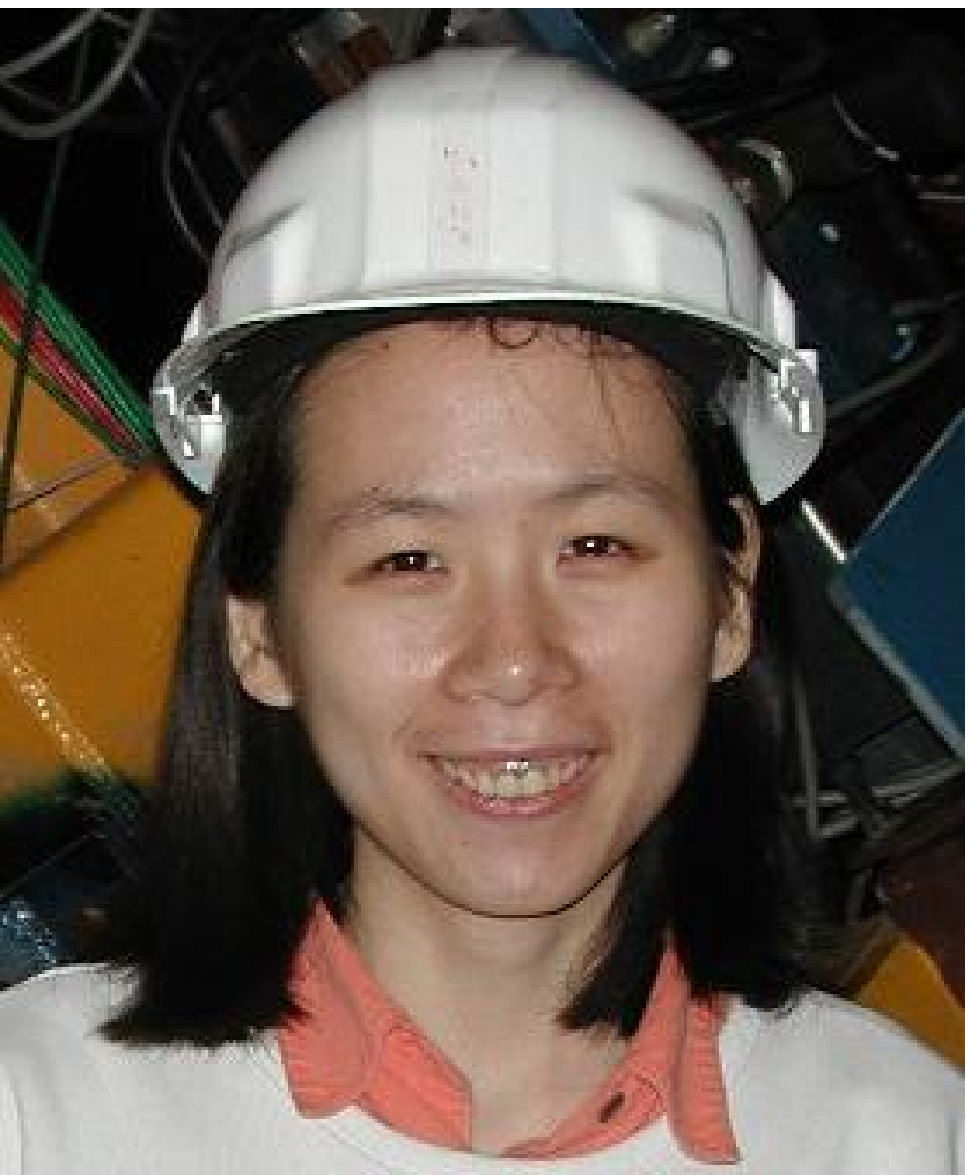,width=35mm}}

\def\Journal#1#2#3#4{{#1} {\bf #2}, #3 (#4)}

\def\NPB{{\em Nucl. Phys.} B}
\def\PLB{{\em Phys. Lett.}  B}
\def\PRL{\em Phys. Rev. Lett.}
\def\PRD{{\em Phys. Rev.} D}
\def\ZPC{{\em Z. Phys.} C}

\def\ra{\rightarrow}

\newcommand{\pbarn} {\ensuremath{\mathrm{pb^{-1}}}}
\newcommand{\fbarn} {\ensuremath{\mathrm{fb^{-1}}}}

\newcommand{\pt}{\ensuremath{p_T}}
\newcommand{\et}{\ensuremath{E_T}}

\newcommand{\gev}{\ensuremath{\rm GeV}}

\newcommand{\gevcsq}{\ensuremath{{\rm GeV}/{\rm c}^{2}}}

\newcommand{\met}{\ensuremath{E_{T}\!\!\!\!\!/}~}

\newcommand{\G}{\ensuremath{\widetilde{G}}}
\newcommand{\GLU}{\ensuremath{\widetilde{g}}}
\newcommand{\SB}{\ensuremath{\widetilde{b}}}
\newcommand{\NONE}{\mbox{$\widetilde{\chi}_1^0$}}
\newcommand{\NTWO}{\mbox{$\widetilde{\chi}_2^0$}}
\newcommand{\CONE}{\mbox{$\widetilde{\chi}_1^{\pm}$}}
\newcommand{\CONEP}{\mbox{$\widetilde{\chi}_1^{+}$}}
\newcommand{\CONEM}{\mbox{$\widetilde{\chi}_1^{-}$}}

\begin{document}
\vspace*{4cm}
\title{Searches for New Physics at the Tevatron in Photon and Jet Final States}

\author{Shin-Shan Eiko Yu}

\address{Fermi National Accelerator Laboratory \\
 Batavia, IL 60510, U.S.A. \\
 for the CDF and D\O\ Collaborations}

\maketitle\abstracts{
  We present the results of searches for non-standard 
model phenomena in photon and jet final states. These searches use data 
from integrated luminosities of 0.7--2.7~\fbarn\ of $p\bar{p}$ collisions 
at $\sqrt{s}=1.96$ TeV, collected with the CDF and D\O\ detectors at the 
Fermilab Tevatron. No significant excess in data has been observed. 
We report limits on the parameters of several models, including: large extra 
dimension, compositeness, leptoquarks, 
 and supersymmetry.}

\section{Introduction}
To date, almost all experimental results have agreed with the predictions 
by the standard model (SM) of particle physics. 
However, several limitations indicate that the SM is not the final theory,
for example: (i) Gravity is not yet described by the SM.
(ii) The electroweak symmetry is broken at energy $\approx 1$~TeV,
 much smaller than the Planck scale $M_{Pl}\approx10^{16}$~TeV (hierarchy 
problem). (iii) The SM does not provide candidates for the dark matter or 
dark energy. In this document, we present the results of searches inspired by 
extensions of the SM: large extra dimension~\cite{ED1}, 
compositeness~\cite{compos}, leptoquarks~\cite{LQ}, 
and supersymmetry (SUSY)~\cite{SUSY,Oscar,GMSB}. 
Specifically, we focus on the searches in final states that contain 
photons ($\gamma$), jets ($j$), or $b$-jets ($b$). 
These searches are based on 0.7--2.7~\fbarn\ of $p\bar{p}$ collisions 
at $\sqrt{s}=1.96$ TeV, recorded with the CDF and D\O\ detectors at the 
Fermilab Tevatron. 
Sections~\ref{sec:LED}--\ref{sec:susy} describe the basic ideas of the 
analysis techniques and present the results of these searches. 
Section~\ref{sec:con} gives the conclusion.

\section{Searches for Large Extra Spatial Dimensions \label{sec:LED}}
In the large extra spatial dimensions model (LED)~\cite{ED1}, SM particles 
are confined to a 4-dimensional membrane and graviton propagates in the 
4+$n_d$ dimension, where $n_d$ stands for the number of additional 
compactified spatial dimensions. The observed Planck scale $M_{pl}$, the 
fundamental Planck scale $M_D$, and the size of the extra dimensions $R$ are 
related by the Gauss Law: \(\left[M_{pl}\right]^2 
= 8\pi R^{n_d}\left[M_D\right]^{n_d+2}\). If $R$ is large compared to the 
Planck length $\approx 1.6\times 10^{-33}$~cm, $M_D$ can be as low as 
1~TeV and effectively solves the hierarchy problem. 
The graviton appears to us, who live in the 4 dimension, like series of 
Kaluza-Klein (KK) states with meV to MeV of mass splittings that can be 
integrated into a massive KK graviton ($G_{KK}$). 
In hadron colliders, we can use two methods to search for indications of 
LED: 
\begin{enumerate} 
\item Look for deviations of the production cross-sections from the SM 
either in absolute values or in shapes, due to exchange of the virtual 
graviton that travels through the extra dimensions. 
The interference and direct gravity terms in the LED cross section are 
parameterized by ${\cal F}/M_S^4$, where $M_S$ is the ultraviolet cutoff of 
the sum over KK states, or the so-called effective Planck scale. The 
formalisms of ${\cal F}$ include: (i) ${\cal F} =1$ (GRW)~\cite{GRW}, (ii)
${\cal F} = \ln(M_S^2/\hat{s})$ for $n_d=2$ and ${\cal F}=2/(n_d-2)$ for 
$n_d>2$, where $\hat{s}$ is the center-of-mass energy of the partonic 
subprocess (HLZ)~\cite{HLZ}, and (iii) ${\cal F}=\pm 2/\pi$ 
(Hewett)~\cite{Hew}. Sections~\ref{sec:angular_ee} and \ref{sec:angular_jj} 
describe this type of LED search using the invariant mass and angular 
distributions of di-electromagnetic (di-EM) and dijet channels, respectively. 
\item Look for emission of real $G_{KK}$ through the production channels 
$q\bar{q}\ra gG_{KK}$, $qg\ra qG_{KK}$, and $q\bar{q}\ra \gamma G_{KK}$, 
with signatures of mono-jet or mono-photon and large $\met$. 
Section~\ref{sec:gmet} describes this type of LED search using the $\gamma\met$ 
final state.
\end{enumerate}

\subsection{Search for LED in the Dielectron and Diphoton Channels
\label{sec:angular_ee}}
\begin{figure}
\begin{center}
\begin{tabular}{cc}
\includegraphics[width=0.5\textwidth]{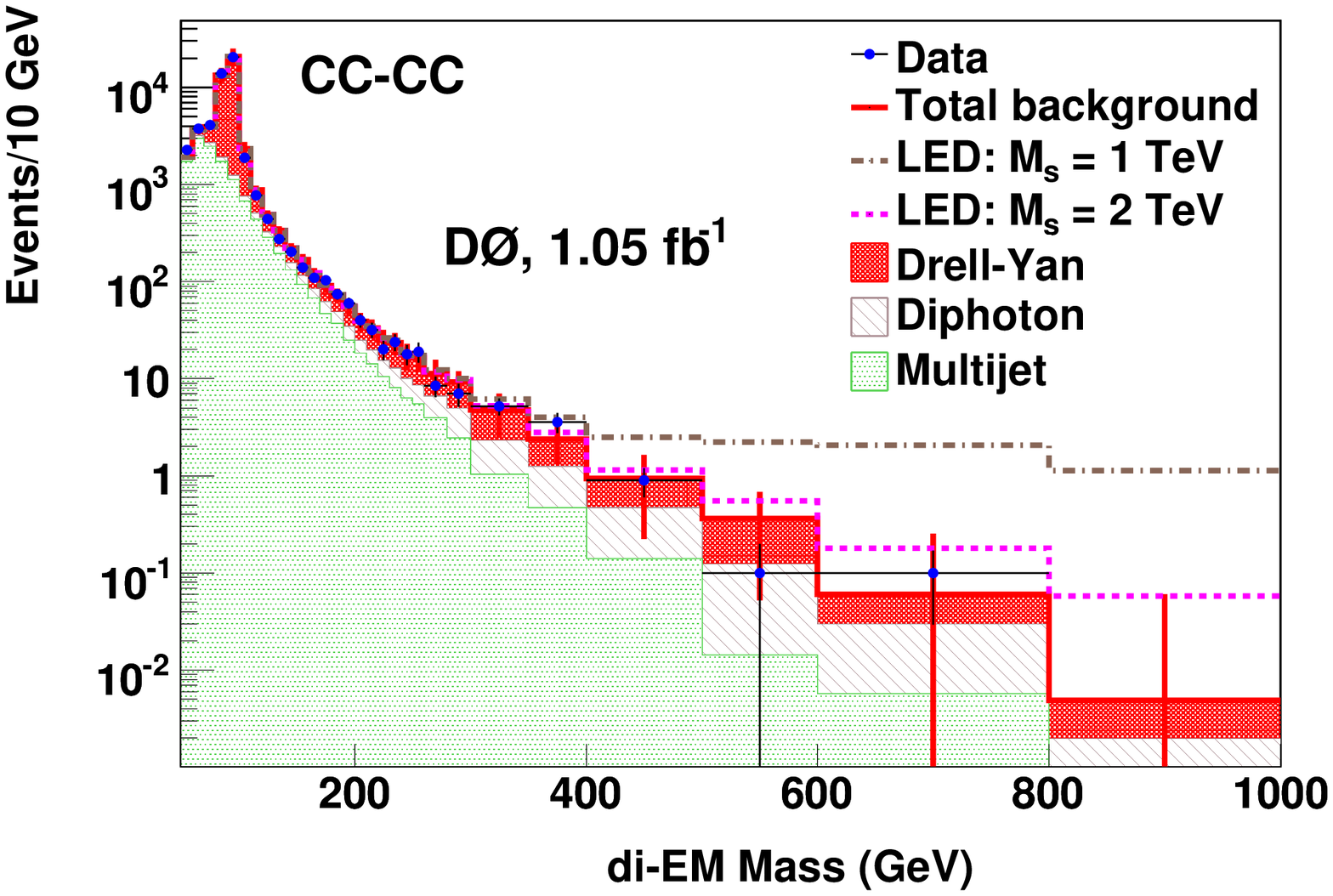} &   
\includegraphics[width=0.5\textwidth]{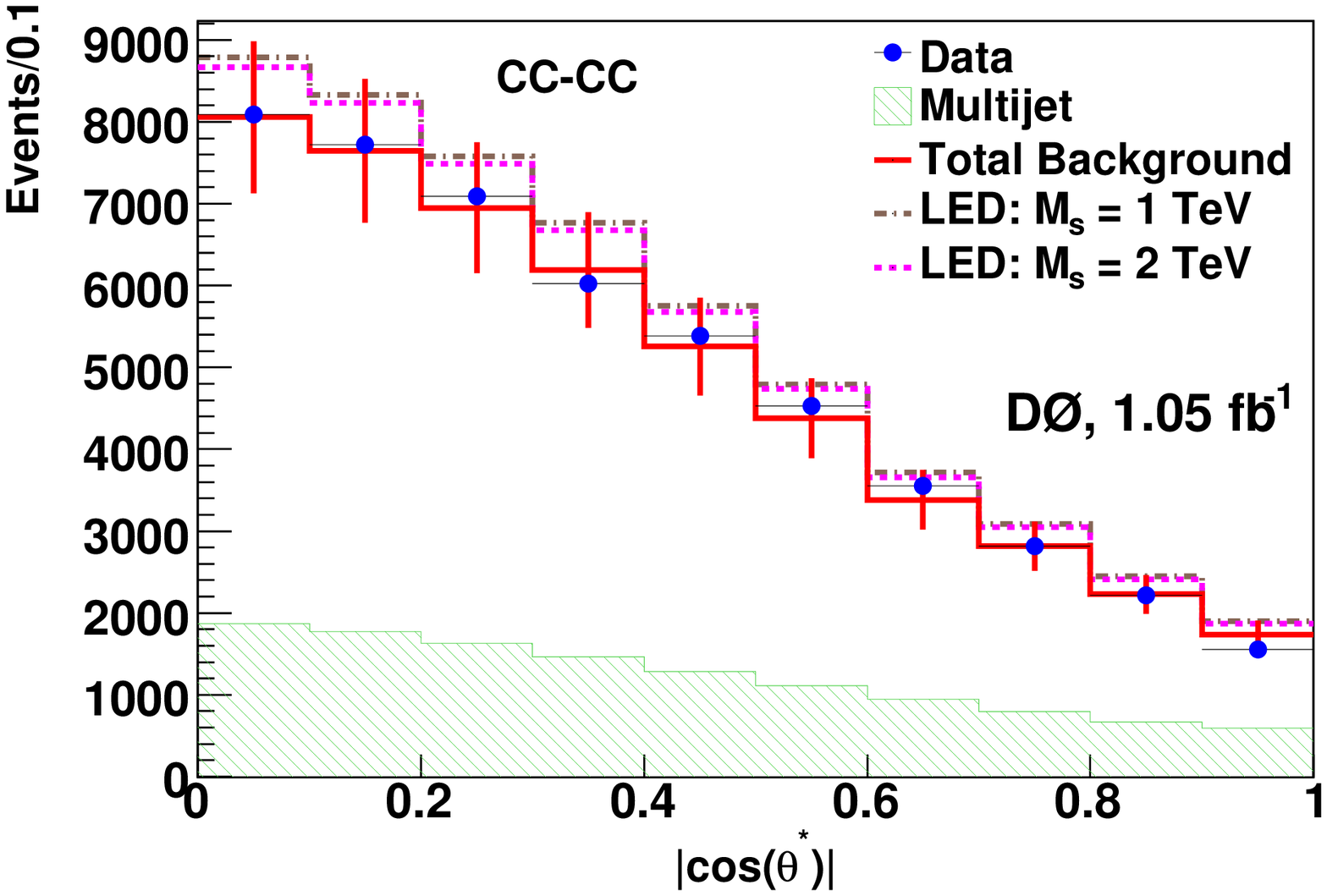} \\
\end{tabular} 
\caption{\label{fig:LEDeegg} 
	The D\O\ LED search:
	the $M_{ee,\gamma\gamma}$ (left) and $\left|\cos\theta^*\right|$ (right)
	distributions, where both EM objects are reconstructed in the central 
	calorimeter. 
	The distributions of the LED signal are obtained by weighting the 
	SM-only full simulation with the ratio of LED+SM to 
	SM parton-level simulations, for $n_d=4$.}
\end{center}
\end{figure}

The D\O\ Collaboration has looked for LED in 1.1~\fbarn\ of $p\bar{p}$ 
collisions, using the two-dimensional distributions of invariant mass 
$M_{ee,\gamma\gamma}$ and angular variable 
$\left|\cos\theta^*\right|$~\footnote{Here, $\cos\theta^*=\tanh(y^*)$, where 
$\pm y^*$ is the rapidity of each EM object in the center-of-mass frame and 
$y^*=\frac{1}{2}(y_1-y_2)$.} 
of two EM objects (combining dielectron and diphoton channels)~\cite{D0eegg}. 
The two EM objects must have $E_T>25$~GeV each,\footnote{We use a cylindrical 
coordinate system in which $\phi$ is the azimuthal angle, $r$ is the radius from 
the nominal beam line, $z$ points in the proton beam direction, and $\theta$ is 
the polar angle measured with respect to the interaction vertex. The 
pseudorapidity $\eta$ is defined as $-\ln(\tan(\theta/2))$. Transverse momentum 
and energy are the respective projections of momentum measured in the tracking 
system and energy measured in the calorimeter system onto the $r-\phi$ plane, and 
are defined as $\pt = p\sin\theta$ and $\et = E\sin\theta$. Missing $\rm \et$ 
($\met$) is defined as the magnitude of the vector $-\sum_{i} E_T^i\hat{n}_i$, 
where $E_T^i$ is the transverse energy deposited in the $i^\mathrm{th}$ 
calorimeter tower for $|\eta| < 3.6$ at CDF and $|\eta| < 4.0$ at D\O, and 
$\hat{n}_i$ is a unit vector perpendicular to the beam axis and pointing at the 
$i^\mathrm{th}$ tower.} and are reconstructed either both in 
the central EM calorimeter ($|\eta| < 1.1$) or one in the central and one 
in the forward EM calorimeters ($1.5 < |\eta| < 2.4$). For the background 
from SM Drell-Yan and diphoton production, the shapes and absolute 
normalizations of their distributions are 
modeled with the {\sc PYTHIA} event generator~\cite{pythia}, followed by a 
D\O\ detector full simulation and a 
mass-dependent $k$-factor ($\sim 1.34$) for the next-to-leading order effect. 
For the QCD background from $\gamma$+jet and multi-jet events, the shapes of 
their spectra are modeled using the data with at least one EM object that 
fails the requirement on 
the shower profile. The normalization of the QCD background is obtained by 
fitting $M_{ee,\gamma\gamma}$ in the range of 
60--140 \gevcsq,  where we expect no LED signal, to a linear combination of 
the SM $ee/\gamma\gamma$ production and QCD background. Then, the fit result is 
extrapolated to the mass region above 140 \gevcsq. 
Figure~\ref{fig:LEDeegg} shows the $M_{ee,\gamma\gamma}$ and 
$\left|\cos\theta^*\right|$ distributions. 
Without observing discrepancy from the background prediction, 
lower limits on $M_S$ are obtained at the 95\%\ confidence level (C.L.): 
1.62~TeV using the GRW formalism, and 2.09--1.29~TeV using the HLZ formalism 
for $n_d=2-7$. These are currently the best limits on $M_S$.


\subsection{Search for LED in the Dijet Channel\label{sec:angular_jj}} 
The D\O\ Collaboration has also used the shape of 
$\chi_\mathrm{dijet}$~\footnote{Here, $\chi_\mathrm{dijet}\equiv (1+\cos\theta^*)/(1-\cos\theta^*)$.} distribution to look for LED in the dijet channel in 
0.7~\fbarn\ of $p\bar{p}$ collisions~\cite{LQ1}. The shape of 
$\chi_\mathrm{dijet}$ is flat for Rutherford scattering, and more 
strongly peaked at small value of $\chi_\mathrm{dijet}$ in the presence of 
LED. Using the shape instead of the absolute distribution makes the search 
less sensitive to the jet energy scale, luminosity, PDF, and renormalization 
scale.  
Jets are reconstructed using the midpoint cone algorithm with 
cone radius of $R=0.7$.\footnote{The $R$ is defined in the $y$ and $\phi$ plane.} 
The four-vectors of jets are corrected 
for the effects of calorimeter response, additional energy from 
multiple $p\bar{p}$ interactions, shifts in $|y|$ due to detector effects, 
and bin-to-bin migration due to finite resolutions. 
Two leading jets are required to have $|y|<2.4$ each, 
invariant mass $M_{jj}>0.25$~TeV$/c^2$, $\chi_\mathrm{dijet}<16$, and 
$\frac{1}{2}|y_1+y_2|<1$. 
The shapes of the corrected $\chi_\mathrm{dijet}$ distributions 
are compared with the SM prediction in bins of $M_{jj}$ from 0.25~TeV/$c^2$ to 
above 1.1~TeV/$c^2$. Since no significant discrepancy is observed 
between the data and SM prediction, limits on $M_S$ are obtained using the 
GRW, HLZ, and Hewett formalisms. However, the limits are not as stringent 
as those from the dielectron and diphoton channels. The same technique 
is also used to set the world's best limits on the compositeness scales 
(see Section~\ref{sec:compos}).

\subsection{Search for LED in the Mono-photon and Large Missing Energy Channel
\label{sec:gmet}}
The CDF and D\O\ Collaborations have searched for LED in 
2.0~\fbarn\ and 2.7~\fbarn\ of $p\bar{p}$ collisions, respectively,
 using events with mono-photon and large $\met$~\cite{CDFLEDgmet,D0LEDgmet}. 
The analyses require one central photon with $E_T > 90$~GeV and 
$\met > 50/70$~GeV for CDF/D\O. Events with extra high $p_T$ tracks or 
jets are removed. The exclusive $\gamma\met$ final state suffers from large 
amount of cosmic rays and beam halos and the analysis would have been 
impossible if an effective rejection was not applied. The CDF analysis 
requires the photon to be in time with a $p\bar{p}$ collision and uses 
topological variables to separate signal from non-collision background, 
such as track multiplicity, angular separation between the photon and the 
closest hit in the muon chamber, and energy deposited in the calorimeters. 
The D\O\ analysis utilizes the transverse and the unique longitudinal 
segmentation of the EM calorimeter. The photon trajectory is 
reconstructed by fitting one measurement in the preshower detector and four in 
the EM calorimeter to a straight line. The $z$ 
position and the transverse impact parameter of the photon, at the point of 
closest approach with respect to the beam line, are required to be within 
10~cm and 4~cm of a $p\bar{p}$ interaction vertex, respectively.\footnote{The 
resolution of the $z$ position is $\approx 3$~cm and the resolution of the 
transverse impact parameter is $\approx 2$~cm.} 
The distribution of the transverse impact parameter is further used to 
estimate the amount of remaining non-collision background. 
After all selections, the dominant background in both analyses is the SM 
$Z\gamma \ra \nu\nu\gamma$ production. Both analyses have not found 
significant excess in data: 
40 observed vs. $46.3\pm3.0$ expected (CDF) and 
51 observed vs. $49.9\pm4.1$ expected (D\O). 
The lower limits on the fundamental Planck scale, $M_D$, are obtained 
at the 95\%\ C.L.: 
1080--900~GeV for $n_d=2-6$ from CDF, and 970--804~GeV for $n_d=2-8$ from 
D\O. The CDF and D\O\ limits using the $\gamma\met$ final state supersede the 
LEP combined limits~\cite{pdg} when $n_d > 3$ and $n_d>4$, respectively. 
The CDF Collaboration further combines the mono-photon+\met\ and 
mono-jet+\met\ channels and excludes $M_D$ below 1400-940~GeV for $n_d=2-6$.

\section{Searches for Quark Compositeness in the Dijet Channel\label{sec:compos}}
The proliferation of quarks and leptons suggests that they may be 
composite structures. The compositeness scale $\Lambda_C$ characterizes the 
physical size of composite states. 
The shapes of $\chi_\mathrm{dijet}$ distributions in bins of $M_{jj}$ as 
described in Section~\ref{sec:angular_jj} are also used to set limits on 
$\Lambda_C$, using the matrix elements in Ref.~\cite{compos}. 
Data with large $M_{jj}$ are more sensitive to large $\Lambda_C$ 
since the deviation from the SM dijet cross section increases as a function of 
$M_{jj}/\Lambda_C$. 
The best lower limits on $\Lambda_C$ have been obtained: 
2.73~TeV for positive and 2.64~TeV for negative interference between the new 
physics and the SM.

\section{Searches for Leptoquarks in the $\ell\ell jj$ and $\ell\met jj$ Channels 
\label{sec:lq}}
\begin{figure}
\begin{center}
\begin{tabular}{cc}
\includegraphics[width=0.5\textwidth]{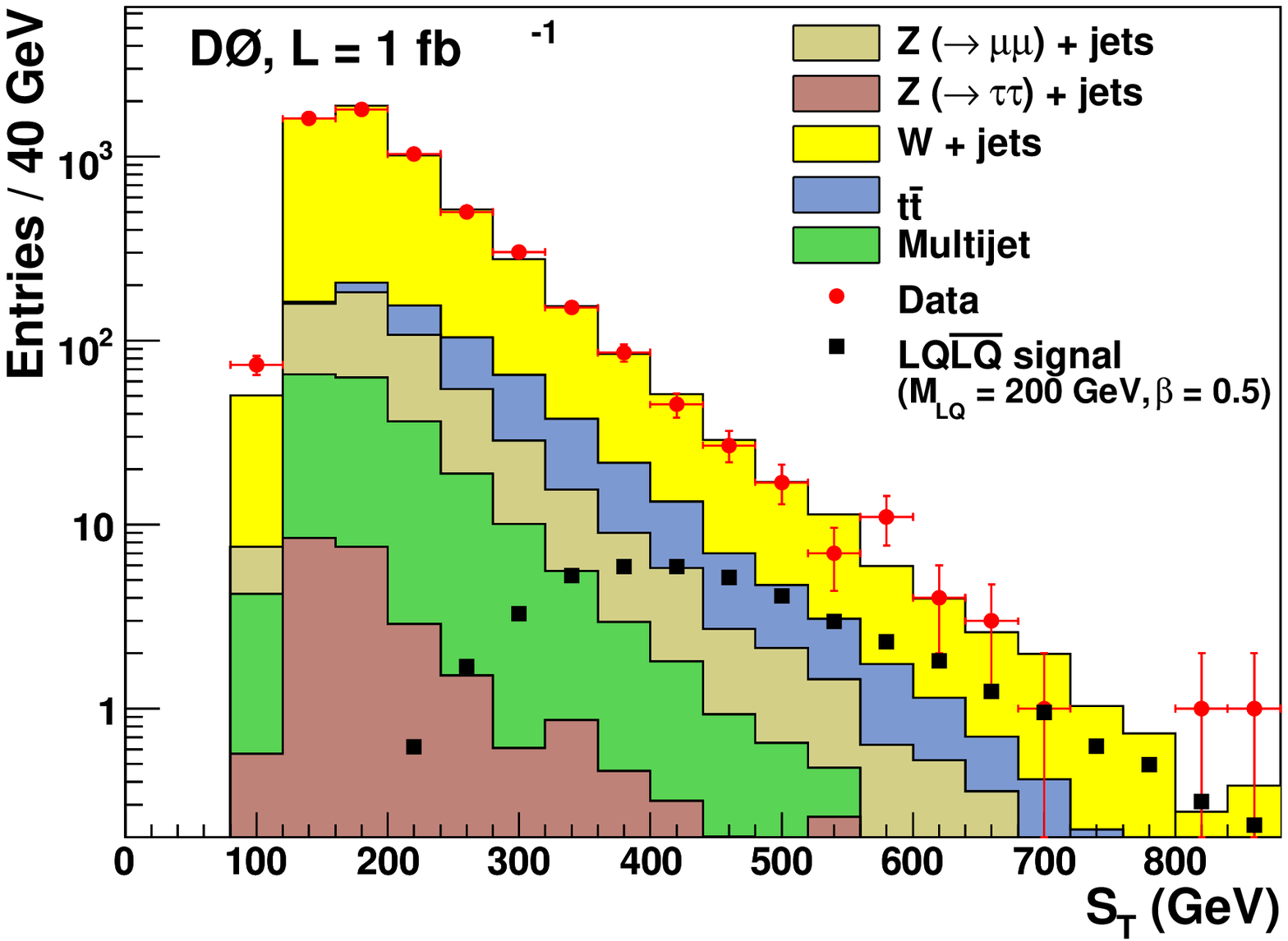} &  
\includegraphics[width=0.45\textwidth]{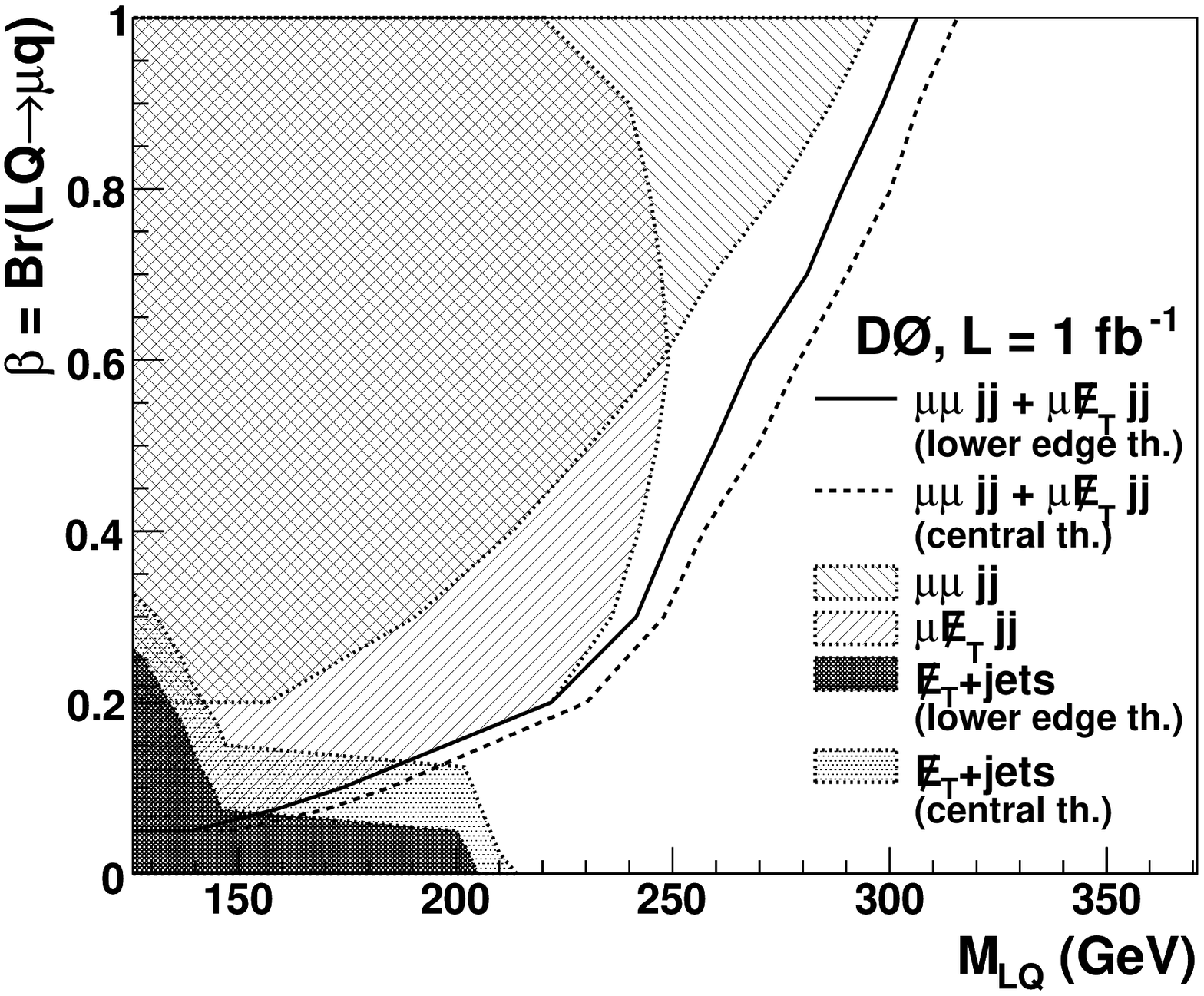} \\
\end{tabular} 
\caption{\label{fig:LQ2} 
	The D\O\ leptoquark search: 
	the $S_T$ distribution of the $\mu\met jj$ events (left) and 
	the exclusion region in the $\beta$ vs. $M_{LQ_2}$ plane (right).}
\end{center}
\end{figure}

Leptoquarks (LQs) are predicted in many models to explain the observed symmetry  
between leptons and quarks, such as technicolor, grand unification theories, 
superstrings, and quark-lepton compositeness~\cite{LQ}. The D\O\ Collaboration 
has looked for pair production of scalar leptoquarks for all three generations 
in 1.0~\fbarn\ of $p\bar{p}$ collisions, assuming LQs couple to 
quarks and leptons within the same generation. The $LQ_1LQ_1\ra eejj$~\cite{LQ1}, 
$LQ_2LQ_2\ra\mu\mu jj$+$\mu\met jj$~\cite{LQ2}, and 
$LQ_3LQ_3\ra\tau\tau bb$~\cite{LQ3} channels are studied, respectively. 
The cross section of pair production depends only on mass of LQ, $M_{LQ}$. 
The coupling of LQ to charge lepton ${\cal B}(LQ\ra \ell q)$ is defined 
as $\beta$ and the coupling to neutral lepton ${\cal B}(LQ\ra \nu q)$ is 
$1-\beta$. Therefore, the final event rates of $\ell\ell jj$ and $\ell\met jj$ 
are proportional to $\beta^2$ and $\beta(1-\beta)$. 
The lepton selections are: 
(i) $eejj$: $E_T^e>25$~GeV, $|\eta_{1,2}^e| < 1.1$ or $|\eta_1^e| < 1.1$ and 
$1.5 <|\eta_2^e| < 2.5$, 
(ii) $\mu\mu jj$ and $\mu\met jj$: $\pt^{\mu}>20$~GeV$/c$, $|\eta_{1,2}^{\mu}| < 2.0$, $\met>30$~GeV, (iii) $\tau\tau bb$: a hadronic and a leptonic 
(decaying to $\mu$) $\tau$ candidate with $\pt>15$~GeV$/c$ each, 
$|\eta_\mathrm{had}| < 3.0$, $|\eta_{\mu}| < 2.0$. 
All jets are reconstructed using the midpoint cone algorithm with 
$R=0.5$ and required to have $E_T^j>25$~GeV and $|\eta_\mathrm{j}|< 2.5$; 
the $\tau\tau bb$ analysis requires at least one jet tagged as $b$-jet.
The variable $S_T$, which is the scalar 
sum \pt\ of the two leptons (either $\ell\ell$ or $\ell\met$), and two 
highest \pt\ jets, is then used as a discriminant to set lower limits on $M_{LQ}$. 
The lower limits on $M_{LQ}$ assuming fixed values of $\beta$ are: 
$M_{LQ_1}^{\beta=1}>292$~\gevcsq, 
$M_{LQ_2}^{\beta=1}>316$~\gevcsq, $M_{LQ_2}^{\beta=0.5}>270$~\gevcsq, and 
$M_{LQ_3}^{\beta=1}>210$~\gevcsq, $M_{LQ_3}^{\beta=0.5}>207$~\gevcsq. 
For the second generation, the $\mu\mu jj$ and $\mu \met jj$ final states are 
also combined to exclude region in the $\beta$ vs. $M_{LQ_2}$ plane. The 
cross-talk of $\mu\mu qq$ in the $\mu\met jj$ events due to the missing muon is 
taken into account. See Figure~\ref{fig:LQ2} for the $S_T$ of $\mu\met jj$ final 
state and the exclusion region in the $\beta-M_{LQ_2}$ plane. 

\section{Searches for Supersymmetry \label{sec:susy}}
Supersymmetry (SUSY) aims to solve the hierarchy problem by introducing 
 superpartners of SM particles~\cite{SUSY}. The spin of SUSY particles 
differs from the original particles by $1/2$. 
For example, the SUSY partners of gluon, graviton, and bottom quark are: gluino 
($\GLU$), gravitino ($\G$), and sbottom quark ($\SB$), and 
carry spin $1/2$, $3/2$, and $0$, respectively. The mixtures of 
SUSY partners of $Z$ boson (zino), photon (photino), and the neutral Higgs 
(higgsino) form four mass eigenstates with spin $1/2$, and are called neutralinos 
($\widetilde{\chi}_{i}^0$, $i=1-4$). 
In the $R$-parity conserving SUSY,\footnote{$R$-parity is defined by the spin 
($j$), baryon number ($B$) and lepton number ($L$): $R\equiv (-1)^{2j+3B+L}$. 
By definition, $R=+1$ for SM particles and $R=-1$ for SUSY particles.} 
the lightest SUSY particle (LSP) is stable and will not decay into SM particles, 
which leaves \met\ and provides possible candidates for dark matter.
Section~\ref{sec:sb} and Section~\ref{sec:GMSB} describe the search for SUSY 
when the LSPs are the lightest neutralino \NONE\ and gravitino \G, 
respectively.

\subsection{Search for Gluino-mediated Sbottom Production \label{sec:sb}} 
In several SUSY models~\cite{Oscar}, sbottom may be light due to the large mixture 
between the left- and right-handed sbottom quarks. If \SB\ is light enough, it may be 
produced via the gluino decay: $\GLU\rightarrow \SB b$. For similar mass, the gluino 
pair-production cross section is an order of magnitude larger than that of sbottom, 
due to gluino's larger color charge and spin. The CDF Collaboration has 
searched for production of gluino-mediated sbottom via the decay chain, 
$\GLU\GLU\rightarrow bb\SB\SB$ and $\SB\rightarrow b\NONE$, which results in a final 
state with 4 $b$-jets and large \met. Event selections are 
at least two jets with $E_T > 25$~GeV (leading jet $E_T>35$~GeV) and $|\eta|<2.4$, 
of which two must be tagged as $b$-jets by the {\sc SECVTX} 
algorithm~\cite{SECVTX}, and $\met > 70$~GeV. Two types of neural network (NN) are 
employed to suppress backgrounds from top pair-production and QCD multi-jet events, 
respectively. 
The requirements on the NN outputs are optimized for two different regions of 
$\Delta m \equiv m(\GLU)-m(\SB)$: (i) small $\Delta m$, $m(\GLU)=335$~\gevcsq\ and 
$m(\SB)=315$~\gevcsq, (ii) large $\Delta m$, $m(\GLU)=335$~\gevcsq\ and 
$m(\SB)=260$~\gevcsq. After these requirements, 2 (5) events are 
observed in data, consistent with the background prediction $2.4\pm0.8$ 
($4.7\pm 1.5$) events for small (large) $\Delta m$ optimization. 
The excluded region on sbottom mass vs. gluino mass from this analysis 
shows a significant improvement to the results from previous 156~\pbarn\ 
analysis and the search for direct pair-production of sbottom 
(see Figure~\ref{fig:susy}).

\subsection{Search for GMSB in the Diphoton and Large Missing Energy Channel 
\label{sec:GMSB}} 
\begin{figure}
\begin{center}
\begin{tabular}{cc}
\includegraphics[width=0.5\textwidth]{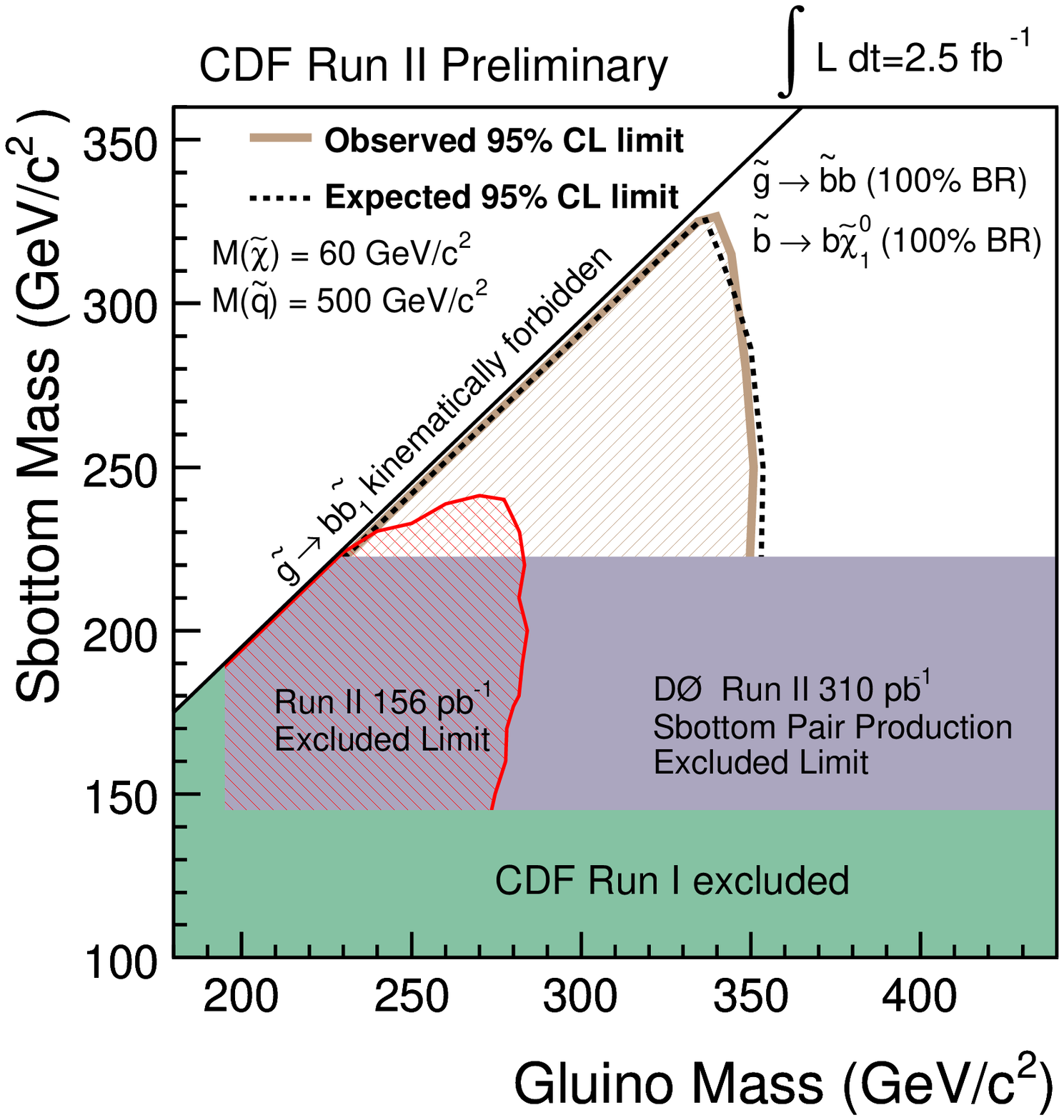} &
\includegraphics[width=0.5\textwidth]{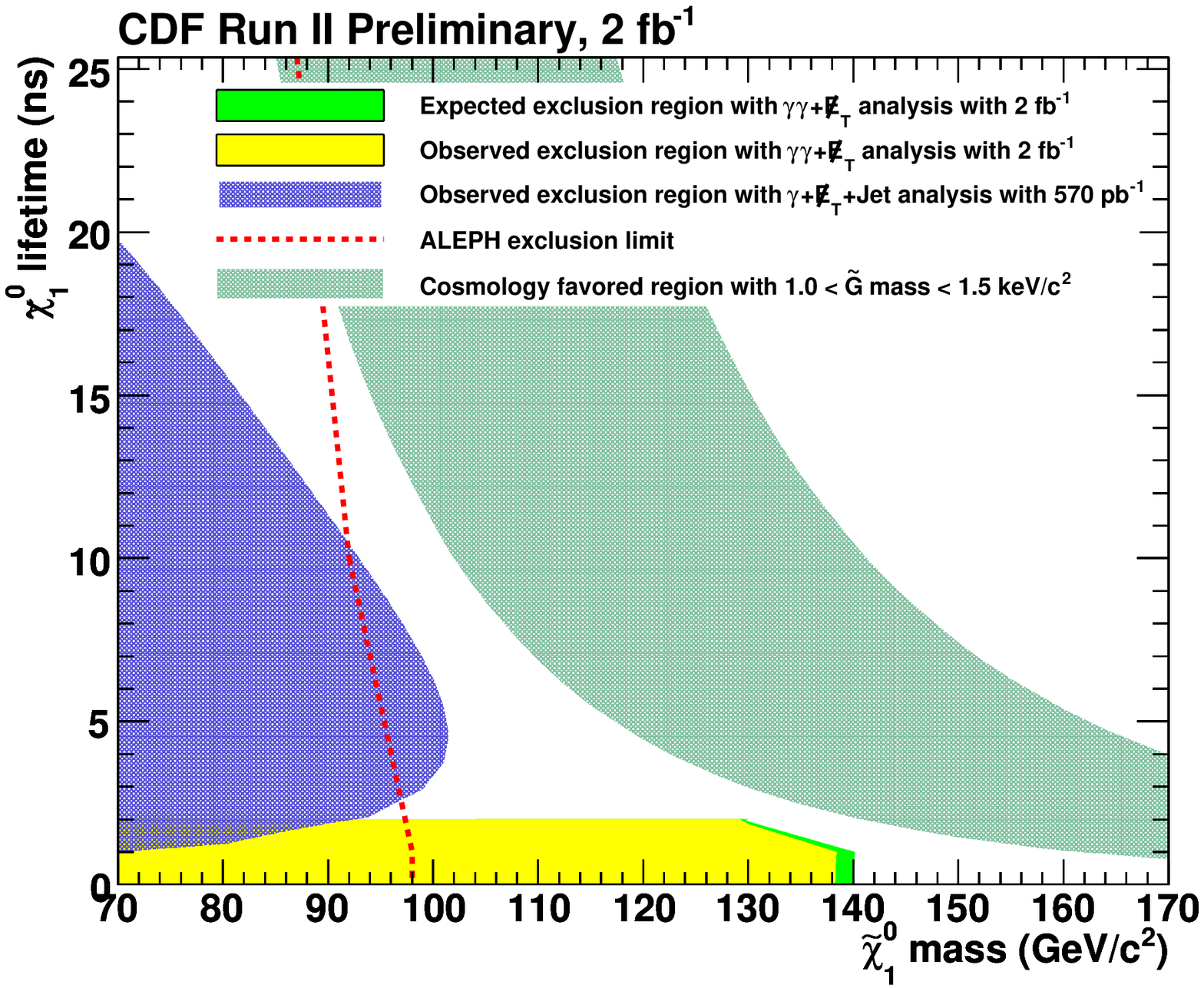} \\
\end{tabular} 
\caption{\label{fig:susy} The CDF gluino-mediated sbottom search (left):
	the exclusion region of sbottom mass vs. gluino mass. 
	The CDF GMSB search (right): 	
	the exclusion region of \NONE\ lifetime vs. \NONE\ mass.
	}
\end{center}
\end{figure}

The CDF Collaboration has searched for gauge-mediated supersymmetry breaking
(GMSB)~\cite{GMSB} in 2.0~\fbarn\ of $p\bar{p}$ collisions. In GMSB, 
the next-to-lightest supersymmetry particle $\NONE$ may decay to the LSP 
\G\ (with a mass of a few keV) via $\NONE\rightarrow\G\gamma$. 
Assuming $R$-parity conservation, pair production of massive SUSY particles, 
such as \NTWO\CONE\ or \CONEP\CONEM, results in a final state with two photons 
and large \met\ due to the escape of the \G\ from the detector. 
This analysis considers a minimal GMSB model (Snowmass Slope SPS 8~\cite{SPS}) 
to quote results as a function of \NONE\ mass and lifetime. 
The requirements are two isolated central photons with $\et > 13$~\gev\ each, 
$\Delta\phi(\gamma_1,\gamma_2)<\pi-0.15$, $H_T > 200$,\footnote{The $H_T$ is the 
scalar sum \pt\ of all identified objects in the events.} and \met\ significance 
$> 3$. The latter three requirements have been optimized to obtain the best 
significance of GMSB signal, and also to reduce the background from $W\gamma$ 
events.\footnote{The electron from $W$ is mis-identified as a photon and the two 
photons are back-to-back due to the large $H_T$ requirement.} 
In order to calculate the \met\ significance, ten pseudo-experiments for 
each event in data are performed. 
The \met\ significance is defined as $-\log({\cal P})$, where ${\cal P}$ is the 
probability for the \met\ drawn from the expected mis-measured \met\ 
distribution\footnote{The expected mis-measured \met\ distribution is modeled by 
studying: (i) the resolution of unclustered energy with zero-jet events in the 
$Z\rightarrow ee$ and fake photon data, (ii) the resolution of jet energy
 with the dijet and $Z$+jet data.} 
to be equal to or larger than the observed \met. 
Further selections are applied to suppress non-collision background 
(cosmics, beam halo, photo-tube spikes). After all selections, 
one event is observed in the data, which is consistent with the background 
prediction, $0.62\pm0.29$ event. Figure~\ref{fig:susy} shows the exclusion region in 
the plane of \NONE\ lifetime (up to 2 ns) vs. \NONE\ mass. For \NONE\ with zero 
lifetime, the mass below 138~\gevcsq\ is excluded at 95\%\ C.L. 
These are the best limits to date. Analysis to search for long-lived \NONE\ 
with more than 2~\fbarn\ is work in progress.

\section{Conclusion \label{sec:con}}
The CDF and D\O\ collaborations have a broad program of searching 
for new physics in photon and jet final states. 
We have not yet found significant excess in 0.7--2.7~\fbarn\ of $p\bar{p}$ 
collisions. We have set the best limits to date on parameters predicted by
 large extra dimension, quark compositeness, leptoquarks, and 
supersymmetry. As more data data are being 
collected, we expect many new and interesting results from both CDF and D\O.

\section*{Acknowledgments}
The author wishes to thank the CDF Exotic and D\O\ New Phenomena group conveners, 
M.~D'onofrio, T.~Wright, T.~Adams, and A.~Duperrin, 
for their suggestions which improved this documentation and 
the presentation in the conference.

\end{document}